\documentclass[twocolumn,showpacs,preprintnumbers,amsmath,amssymb]{revtex4}

\usepackage{graphicx}
\usepackage{dcolumn}
\usepackage{bm}

\begin{document}
\title{Properties of N=90 Isotones within the Mean Field Perspective}%
\newcommand{\nucl}[2]{\ensuremath{^{#1}}\mbox{#2}}
\author{E.~Ganio\u{g}lu$^1$\footnote{Email address: 
ganioglu@istanbul.edu.tr}, R.~Wyss$^2$, 
P.~Magierski$^3$\\}

 \affiliation{%
$^1$ Science Faculty, Physics Department, Istanbul University, TR-34459, 
Istanbul,Turkey.\\
$^2$ Department of Physics, KTH (Royal Institute of Technology), SE-10691 
Stockholm, Sweden \\
$^3$ Faculty of Physics, Warsaw University of Technology, ul. Koszykowa 
75, 00-662 Warsaw, Poland \\
}%

\date{\today}

\begin{abstract}
In recent years, the N=90 isotones have been investigated to a large 
extent in relation to studies of quantum phase transitions. In this paper, we use the
mean field approach with pairing-deformation self consistent 
Total Routhian Surface (TRS) calculations to study the
N=90 isotones and neighbouring nuclei. The important probes, such as moments of inertia, 
quadrupole moments, the energy ratio of E($4^+_1$)/E($2^+_1$), octupole 
and hexadecapole degrees of freedom are considered and the calculated results are compared with the available 
experimental data. From a microscopic point of view, 
the N=90 isotones characterize the onset of the deformed region and are very well described by mean field calculations.
The results are compared with those from other studies in beyond mean-field approximations. Shape coexistence phenomena in the region of interest 
are discussed.

\end{abstract}

\pacs{21.60.-n, 31.50.-x, 27.70.+q, 29.30.Kv}
\maketitle

\section{Introduction}
Many physical systems undergo phase transitions. Despite the fact that 
atomic nuclei are finite systems they too exhibit phase transitions like in their 
shape and these changes markedly modify the properties of the entire system. 
Following the introduction by Iachello~\cite{Iachello-1,Iachello-2} of 
a simple model of critical point symmetries of the shape transitions X(5) 
from spherical vibrator to axial rotor and E(5) from spherical to gamma-
unstable 
nuclei, there has been considerable effort invested in both theoretical 
and 
experimental studies of these dynamical symmetries. The development of 
collectivity as one moves away from closed shells in nuclei is a topic of 
abiding 
interest and Iachello's introduction of the X(5) critical point symmetry 
points 
the way to where one might find the critical point in such quantum phase 
transitions, see also reviews~\cite{Cjenar,Bonatsos-1,Bonatsos-2,Fortunato}. 

Phase transitions have also 
been studied from a different perspective, using a variety of different 
mean field methods like 
relativistic Hartree-Bogoliubov 
theory~\cite{bonatsos,Rodrigues,Niksic,Tomas}. In some of these works,
 it was suggested that the phase transition can be related to a prolate-oblate 
shape transition
~\cite{bonatsos,Sarriguren}. In the present work we will also examine the 
issue of 
phase transitions in N=90 isotones from a mean field perspective.
In this work, triaxial deformations are taken into account, in order to 
describe the possible 
prolate to oblate shape transitions adequately if they exist. This 
ensures that any 
spurious minima corresponding to saddle points in the potential energy 
landscape, are found. 
Our results reveal that the oblate minima in the N=90 region are spurious 
and that 
there is no evidence for prolate oblate shape co-existence in this mass 
region. This result, showing that the oblate minima are saddle points in $\beta-\gamma$ plane 
is consistent with the results of Ref.~\cite{Tomas,Li}

In addition to the potential energy surfaces, the moments-of-inertia and 
quadrupole moments 
were calculated and compared to the experimental values. Since the issue 
of quantum phase 
transitions in a finite system is of particular interest we believe that a better 
understanding 
of these three indicators will provide deeper insight into this 
phenomenon. It is an open issue how 
nuclei in which these observables are well described by the
mean field approach relate to phase transitional 
behaviour and/or
reveal a critical point symmetry. 
It is well known that the cranking 
moment-of-inertia 
describes the structure of  low and high spin yrast states rather well, 
provided 
the mean field has a well defined, deformed shape. Thus, it is of 
high interest to determine the ratio  $R(4/2)$ for the N=90 isotones in 
realistic calculations. 
This will also allow us to establish the region in which the deformed mean field
approximation  is valid, i.e. the region of deformed nuclei within the mean field concept. 

In the next section our calculations are described and the results are discussed. A
comparison is then made with the experimental data. 

The conclusions are given in section III.

\section{The Results}

The deformed mean field for our investigation
is based on a deformed Woods-Saxon potential~\cite{cwiok} and the 
Strutinsky Shell
correction approach~\cite{Strutinsky1,Strutinsky2}. Rotational states are 
generated by  means of the cranking approximation, that is well suited to application 
to deformed nuclei. Pairing correlations are included via a seniority pairing 
force and a double-stretched quadrupole pairing interaction. The time-odd 
component of the latter is of particular importance if we are to obtain a correct 
description of the moment-of-inertia ~\cite{wyss-satula,satula-wyss}. The pairing 
Hamiltonian is calculated in a self-consistent fashion at each frequency and 
each deformation point. In order to avoid the spurious break down of the 
pairing field, approximate particle number projection via the Lipkin Nogami
method is employed. For further details
of the method, we refer the reader to~\cite{NPA-578,PST-56}. The total 
energy is minimized with respect to the shape parameters ~\cite{wyss}.
The method has successfully been applied over the entire nuclear chart, 
giving in general a very accurate description of the
rotational spectra of deformed nuclei, see e.g.  
~\cite{PRC-56,PRL-83,PRL-87,PRC-60}
The accuracy with which the TRS calculations reproduce the properties of high spin 
rotational bands gives us some confidence in the method when it is employed in 
describing the properties of deformed nuclei more generally with the mean field.

\begin{figure}[htbp]
\rotatebox{0}{\scalebox{0.37}{\includegraphics[angle=270]{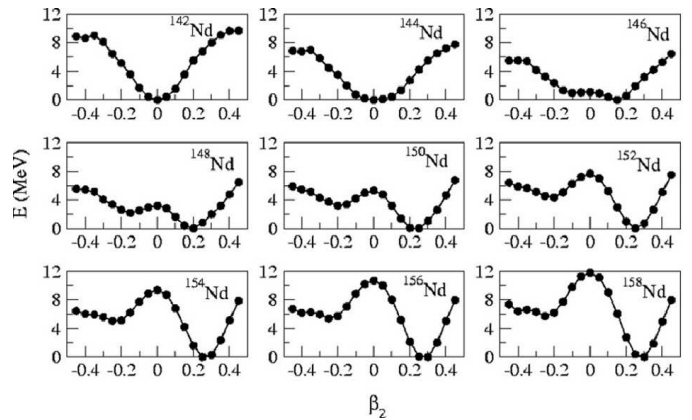}}}
\caption{ Potential energy as a function of $\beta_2$ for the even-even 
Nd nuclei from N=82 - 98.}
\label{fig:level1}
\end{figure}

We start by examining the potential energy surfaces as they emerges from 
calculations for axially symmetric shapes for the chain of 
even-even 
Nd isotopes, from N=82 to N=98, (see Fig.~\ref{fig:level1}). 
The potential energy curves in the Nd-isotopes as a function of N 
might serve as a textbook example for the onset of deformation, see e.g.~\cite{Naz}. 
The N=82 isotope is spherical and, one expects a vibrational excitation 
structure. 
On the other hand for N=98 we see a well developed deformed shape, 
having 
$\beta_2\approx 0.28$ and hence rotational structure. At N=84, the equilibrium shape is still spherical, 
but considerably softer. For N=86, 
there is effectively a broad minimum
and at N=88 a deformed shape emerge.
The deformation increases smoothly with increasing neutron number up to 
$^{150}Nd$ and there is little change above $^{150}Nd$. 
The mean field approximation is
applicable, when the fluctuations in $\beta$ are smaller than the mean 
field value, 
which is 
equivalent to stating that the zero-point motion is confined within the 
barriers of the 
potential surface. For those cases the cranking approximation is well justified. 
Our results indicate, that for N=90, this indeed is the case.

\begin{figure}[htbp]
\rotatebox{0}{\scalebox{0.43}{\includegraphics{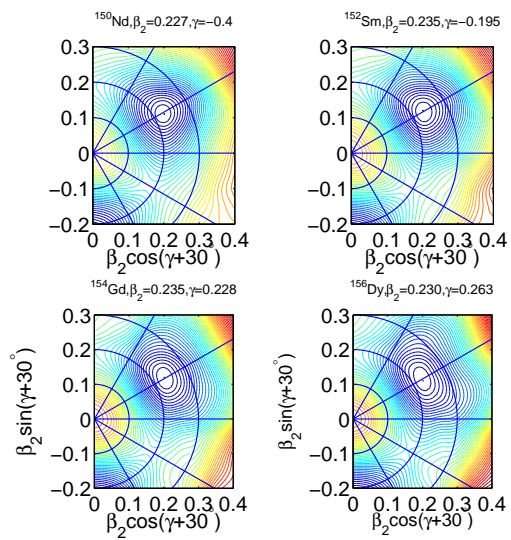}}}
\caption{ Potential energy surfaces for the N=90 isotones, 
$^{150}Nd$,$^{152}Sm$,$^{154}Gd$,$^{156}Dy$. Energy difference between adjacent contour lines is 100 keV.}
\label{fig:level2}
\end{figure}

\begin{figure}[htbp]
\rotatebox{0}{\scalebox{0.35}{\includegraphics[bbllx=0 ,bblly=70 ,
  bburx=595 ,bbury=600, clip=]{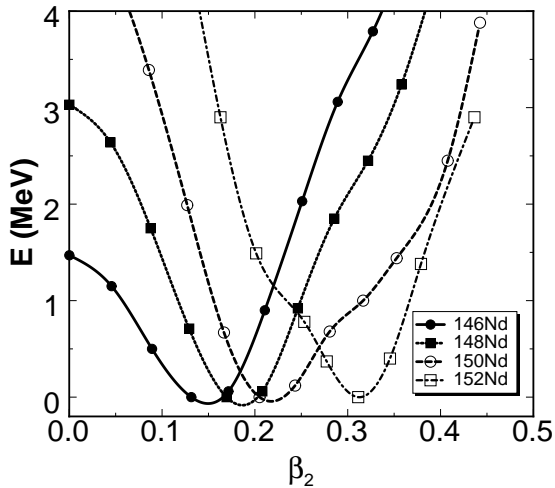}}}
\caption{Potential energy as a function of $\beta_2$ for some even-even Nd 
isotopes.}
\label{fig:level3}
\end{figure}

From the potential energy curves shown in Fig.~\ref{fig:level1} one may 
deduce that there is 
oblate-prolate shape coexistence in some of the Nd-isotopes. Relaxing the condition 
of axially symmetric
 shapes, and allowing triaxiality to play a role one realizes that the 
oblate 'minimum' is actually 
a saddle point and that there is no sign of shape coexistence (see 
Fig.~\ref{fig:level2}).
Clearly, any conclusion about prolate-oblate shape co-existence should be 
made with care when the symmetry is 
restricted to axially symmetric shapes. Oblate minima in the N=90 region 
are an artefact of 
the symmetry restriction in the model.
The calculations show that the N=90 isotones from Nd to Dy all have well 
developed minima at $\beta_2 \approx 0.23$.
In order to address the question of whether a simple mean field model can give a 
quantitative description of nuclei at or near a critical point, we compare 
our calculations with the Skyrme Hartree-Fock BCS
(HFBCS) calculations described in Ref.~\cite{HFBCS}. Qualitatively, the 
calculations are in agreement, see
Fig.~\ref{fig:level3}. The 
potential depths of the energy curves are very similar, revealing that at 
N=86, the nuclear shape is becoming deformed with a similar value of 
the deformation parameter for all the values of Z. 
The main difference in the calculations
is related to the more sudden increase in deformation between N=90 and 
N=92 in the Skyrme HFB 
calculations, whereas in the calculations with Woods Saxon potential, the deformation 
increases smoothly with 
neutron number. This difference most probably reflects the difference in 
the single particle spectrum, 
related to deformed shell gaps below/above the [660]1/2 Nilsson orbit. 
It is also interesting to compare our results with the Hartree–Fock–Bogoliubov (HFB)
 approach using the Gogny interaction (Ref.~\cite{Robledo}).
As one would expect this study gives potential energy curves in the region of interest similar to ours.
They found that the transitional behaviour appears for the N=86-90 isotopes, $^{146-150}Nd$. 
The main difference is that they find a wide minimum on the prolate side as well as an additional minimum on the oblate side. As indicated earlier the suggestion of shape co-existence should be considered with care when restricting calculations to axial symmetric shapes.

As discussed above, the potential energy surfaces reveal nicely the 
calculated equilibrium shapes
as a function of deformation, but do not allow further 
conclusions to be made on the applicability
of the model with respect to observables such as the quadrupole moment 
and moments-of-inertia. 
In the deformed mean field picture, the transitional quadrupole moment is 
an excellent 
observable to characterise the shape of the nucleus. In the collective 
model of Bohr and Mottelson,
assuming a uniformly rotating body with given spins, the transitional 
moment is given by
\begin{equation}
B(E2;KI_{1}\rightarrow
KI_{2})=\frac{5}{16\pi}e^{2}Q_{0}^{2}<I_{1}K20|I_{2}K>^{2}
\end{equation}
where the intrinsic quadrupole moment $Q_\circ$ is obtained as the 
integral of the charge distribution.
One can directly use the deformation parameters of the 
potential to deduce the 
corresponding quadrupole moment. Instead, in the present calculations, we 
calculate the expectation 
values of the quadrupole operators $Q_{20}$ and  $Q_{22}$ 
microscopically from
the occupation probability of the single particle levels of the Woods-Saxon potential:
\begin{equation}
<Q_{2\mu}> = Tr< q_{2\mu} \rho(\omega)>,
\end{equation}
where $ q_{2\mu} $ is the matrix of the single particle quadrupole 
moments, and $\rho(\omega)$
the density matrix at rotational frequency $\omega$ as
obtained in the self-consistent HFB diagonalisation.
\begin{figure}[htbp]
\rotatebox{0}{\scalebox{0.4}{\includegraphics[bbllx=0, bblly=10,
  bburx=600, bbury=500, clip=]{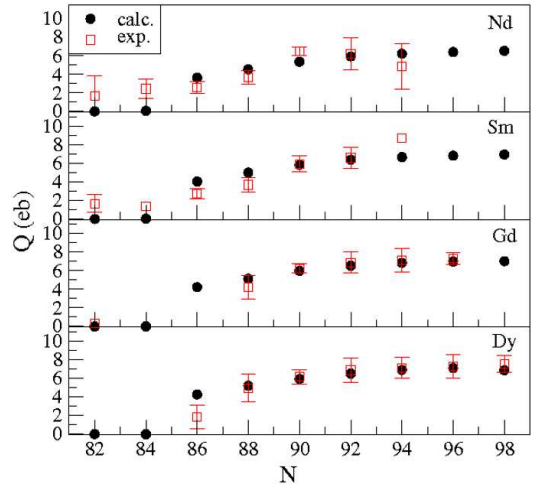}}}
\caption{ The experimental(squares)~\cite{nds} and calculated(circles)
Quadrupole moments for even-even Nd, Sm, Gd and Dy isotopes 
as a function of neutron number N.} \label{fig:level4}
\end{figure}
Ref~\cite{Niksic} asserts that if one is to recognise a critical point symmetry 
in a particular nucleus B(E2) ratios are one of two properties that must be calculated.
Although B(E2) rates are not possible to obtain directly from the mean-field calculations, we compare our calculated quadrupole moments to the experimentally deduced ones,
using the relation between experimental B(E2) ratios and quadrupole moments, Eq.~1.
For obvious reasons, the deformed mean field does not give a quadrupole 
moment when the deformation is zero.
In contrast, experiment reveals a sizeable moment, 
showing that the deformed mean field model does not apply to those nuclei
Fig.~\ref{fig:level4}(For the experimental data see~\cite{nds}).
The transition probability 
can be calculated in e.g. the Random Phase approximation (RPA), as the 
first order extension
of the mean field. Indeed, RPA calculations nicely depict the
drop in excitation energy from N=82 to N=84~\cite{ela}. 
In our calculations, we find a sudden onset of the quadrupole moment at 
N=86, related to
the fact that the deformation has a non zero value. This does not 
necessarily imply that
the approximations underlying the calculations are valid. In experiment 
one notices a smooth increase in 
the transitional quadrupole moment, revealing the transition from 
vibration-like to rotational structure.

\begin{figure}[htbp]
\rotatebox{0}{\scalebox{0.44}{\includegraphics[bbllx=50, bblly=70,
  bburx=650, bbury=600, clip=]{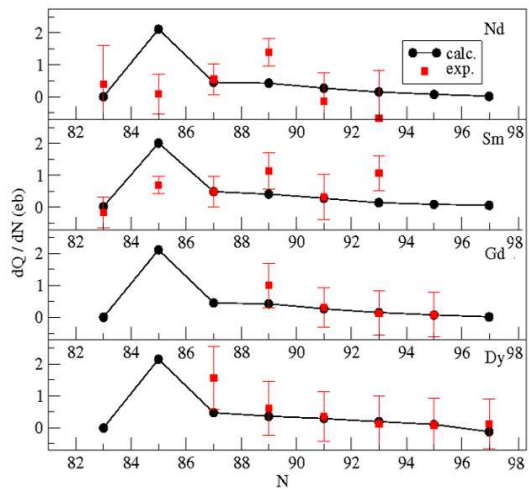}}}
\caption{ The change in Quadrupole moments for experiment
(squares)~\cite{nds} and calculations(circles) as a function of neutron 
number in the even-even Nd, Sm, Gd and Dy nuclei for N=82-98 nuclei from top to bottom respectively.}
\label{fig:level5}
\end{figure}

The change in the quadrupole moment with N, dq/dN is expected 
to be largest at the point of transition from vibrational like to deformed.
In the chains of nuclei discussed here from Nd to Dy, it occurs 
between N=88 and N=90. This result is consistent with the result of 
the study of the beyond mean field approximation using the Gogny interaction 
given in Ref.~\cite{Tomas}. 
(see Fig. ~\ref{fig:level5})
For transitional nuclei the contribution to the quadrupole
transition originates from a superposition of uniform rotational
and vibrational motion. Since the rotational component is larger in size, 
one expects that the calculations assuming uniform rotation may yield reasonable 
results, even when the structure of the first excited 
state still has an appreciable component of vibrational motion. Indeed, 
the calculated quadrupole moments
agree rather well, starting from N=86 and certainly by N=88.

\begin{figure}[htbp]
\rotatebox{0}{\scalebox{0.45}{\includegraphics[bbllx=60, bblly=80,
bburx=650, bbury=600, clip=]{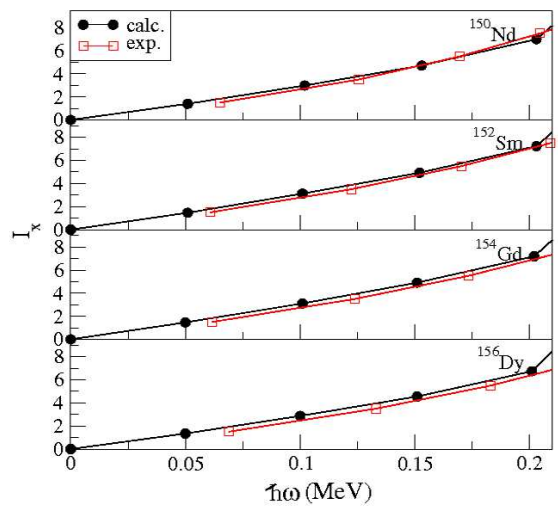}}}
\caption{ The experimental(squares)~\cite{nds} and calculated(circles) 
spin($I_{x}$) 
versus rotational frequency ($\hbar\omega$) in the even-even 
Nd, Sm, Gd and Dy nuclei for N=82-98 nuclei from top to bottom respectively.}
\label{fig:level6}
\end{figure}

The calculated spins as a function of the rotational frequency, from 
which one can deduce the
moment-of-inertia, are another sensitive probe of the validity of the cranking 
approximation.
The moments-of-inertia are described rather well for deformed nuclei, for 
which the mean field
approximation is valid. In Fig.~\ref{fig:level6}, we compare the 
calculated spins to experiment for 
all of the N=90 isotones discussed in our study. 
Clearly,
the calculations agree very well with the experimental data in the low 
spin regime, which
is of relevance to our discussion.
This indicates that the N=90 isotones are well described by
the mean field approach, indicating the validity of a static deformed mean field. 

\begin{figure}[htbp]
\rotatebox{270}{\scalebox{0.37}{\includegraphics[bbllx=60, bblly=70,
  bburx=600, bbury=680, clip=]{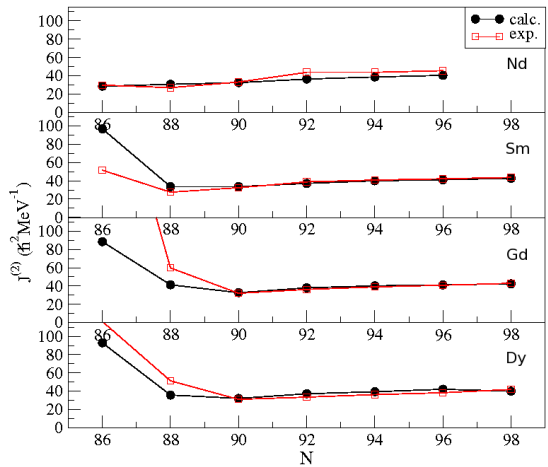}}}
\caption{ The experimental(squares)~\cite{nds} and calculated(circles) 
dynamical moments-of-inertia.
For the even-even Nd, Sm, Gd and Dy isotopes as a function of neutron 
number N}
\label{fig:level7}
\end{figure}

The most 
sensitive probe
in the comparison with experiment is the second moment-of-inertia, $J^2$.
We used the calculated spin alignment and frequency values to calculate 
the second moment-of-inertia 
by means of the well known equation $J^2=dI/d\omega$ and the experimental 
moments-of-inertia were 
obtained from the following equation.
\begin{equation}
J^2=\frac{I_x(\omega_2)-I_x(\omega_1)}{\hbar(\omega_2-\omega_1)}
\end{equation}
For an ideal vibrator, the  $J^2$ moment-of-inertia is expected to be 
infinite. It will be very
large for anharmonic vibrators. On the other hand, in a regime of 
deformed nuclei, we expect a
smooth increase in $J^2$ with increasing deformation. Hence, studying the 
transition from vibrator to rotor
in an isotopic chain,
should reveal a minimum in the moment-of-inertia as a function of neutron 
number.
Indeed, comparing the evolution of the moments-of-inertia in the 
different isotopic chains from Nd to Dy,
it reaches a minimum at N=90 and then smoothly and steadily increases for 
larger N values 
which is seen in Fig.~\ref{fig:level7}. Starting at N=90, the
moment-of-inertia increases smoothly, with no sign of sudden changes. The 
calculated second
moment-of-inertia agree very well with experiment from N=90 onwards and 
show the same behaviour.
The reason for the larger moments-of-inertia in some of the N=88 isotones 
reflects the fact that in 
these nuclei at these frequencies, an alignment is taking place, i.e. 
there is a sizeable
 single particle contribution to the
moment. At lower frequencies, we obtain a lower moment-of-inertia in N=88 
as compared to N=90, indicating
that the cranking approximation has limited validity when it comes to the 
N=88 isotones.

\begin{figure}[htbp]
\rotatebox{0}{\scalebox{0.65}{\includegraphics{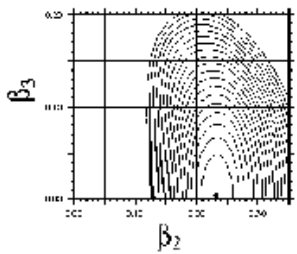}}}
\caption{Calculated octupole deformation versus quadrupole 
deformation for $^{150}Nd$. Energy difference between adjacent contour lines is 100 keV.}
\label{fig:level}
\end{figure}

\begin{figure}[htbp]
\rotatebox{0}{\scalebox{0.4}{\includegraphics[bbllx=0, bblly=10,
  bburx=300, bbury=300, clip=]{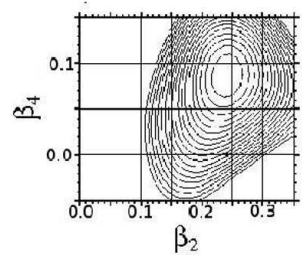}}}
\caption{Calculated hexadecapole deformation versus quadrupole 
deformation $^{150}Nd$. Energy difference between adjacent contour lines is 100 keV.}
\label{fig:level8}
\end{figure}

\begin{figure}[htbp]
\rotatebox{270}{\scalebox{0.35}{\includegraphics[bbllx=200, bblly=130, 
bburx=650, bbury=950, clip=]{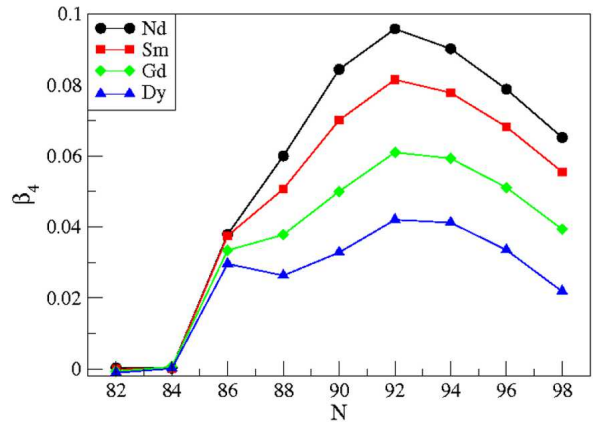}}}
\caption{ The calculated hexadecapole deformations of the even-even 
Nd,Sm,Gd and Dy 
isotopes as a function of neutron number N.}
\label{fig:level9}
\end{figure}

In order to investigate the dependence on other shape parameters, we calculated potential energy surfaces for octupole and hexadecapole deformations.
For the octupole degree of freedom, one finds considerable softness that is largest at 
N=86,88 (~0.8-0.9)for the Nd and Sm isotopes while it occurs at N=84 
(~0.5) for the Gd and Dy isotopes. 
In the case of the N=90 isotones the octupole softness decreases with 
increasing proton number which indicates that N=90 is not a major point 
of shape changes 
in the octupole direction~\cite{witek}. (See Fig.~\ref{fig:level}).
Moreover as a result of our calculations we found sizeable $\beta_4$ 
values for
 N=90 and N=92~\cite{ela-AIP} See Fig.~\ref{fig:level8} and Fig.~\ref{fig:level9}.

The energy ratio $R_{4/2}$ 
is another important indicator of structure. We used an averaging procedure in our calculations in order to determine the E(4)/E(2) ratio . 
First the mean field moments of inertia are calculated microscopically
at each frequency.
Then using the relationship between energy and moments-of-inertia
we calculate the $R_{4/2}$ 
ratio in rotational
basis using the following formula:
\begin{equation}
E_{rot}=\frac{\hbar^2}{2J}I(I+1)
\end{equation}
and
\begin{equation}
E(4^+_1)/E(2^+_1)=3,33\frac{J^{(2)}}{J^{(4)}},
\end{equation}
where $J^{(2(4))}$ is calculated at the frequency for the 
$2(4)^+\rightarrow 0(2)^+$ transition, using interpolation.
The calculated ratios are shown in Table~\ref{table:gam}. They clearly 
show that
the mean field estimates are consistent with the experimental
data with respect to this energy ratio. The experimental energies are 
taken from~\cite{nndc}
\begin{table}[htbp] \caption{\label{table:gam}The calculated and 
experimental E($4^+_1$)/E($2^+_1$) 
ratio for the N=90 isotones.}
\begin{ruledtabular}
\begin{tabular}{l r r r}
Nucleus & Calculated & Experimental \\ & E($4^+_1$)/E($2^+_1$) & 
E($4^+_1$)/E($2^+_1$) \\

\hline\\
 $^{150}Nd$&2.90  &2.927&  \\
 $^{152}Sm$&3.04   & 3.009& \\
 $^{154}Gd$&3.12   & 3.015& \\
 $^{156}Dy$&3.07  & 2.934& \\
\end{tabular}
\end{ruledtabular}
\end{table}

The moment-of-inertia is changing with frequency, mainly due to the 
change in the pairing field. This effect
is not negligible, particularly for the N=90 isotones.
Apparently,  N=90 nuclei are nicely described within the cranking 
approximation.
A value of 2.9 appears to be a lower limit for the $R_{4/2}$ 
ratio that can be obtained in cranking calculations. 
The value of 3.3 obtained at the symmetry limit of SU(3) corresponds to
an ideal rigid rotor. Nuclei are not rigid rotors and the change in the moment-
of-inertia, as obtained from
the cranking calculations, yields a rather accurate  value of the response 
of the nuclear mean field.
Pairing correlations are of course crucial in this context and are the single most 
important factor causing the increase 
in the moment-of-inertia with increasing spin. 
The dynamic change of pairing correlations and their influence on a possible 
phase transition is not considered in algebraic approaches. Therefore one 
expects
a 'smearing effect', due to fluctuations in the nuclear wave function. 
It implies that the value of 2.91, as obtained for the critical point 
symmetry has to
be treated with caution in any comparison with real nuclei. 
As stated in Ref.~\cite{casten-1}, nucleon numbers only take discrete 
values and there is no continuous parameter
associated with them. From our calculations it emerges that N=90 is well 
described within the mean 
field model  and N=88 and N=90 mark the borderline between nuclei that can 
be described  
well within the mean field and those that reveal deficiencies. The 
presence of pairing correlations
makes it difficult to use 2.91 as a number that definitively 
characterizes the phase transition in nuclei.
One may assume that the value is smaller in real nuclei. 
In addition, the presence of pairing correlations in the ground state
 results in two major changes:
i) the critical value will be shifted to a lower neutron number and ii) 
the phase transition is most likely to be 
smeared out due to the presence of strong fluctuations.

\section{Conclusion}
The present paper investigates the validity of the deformed mean field for the description of N=90 
isotones with pairing-deformation self consistent 
Total Routhian Surface (TRS) calculations. 
In order to understand how the nuclear shape evolves the calculations are carried out in the range of 
N=82-98. 
Our results show that the two minima appearing in axially symmetric calculations correspond to 
a spurious one at oblate shape and a proper one at prolate shape. 
We compared our calculations with the Skyrme Hartree-Fock BCS(HFBCS) calculations  and 
to previous studies in the beyond mean field approximation.
Previous proposals of shape co-existence are not
verified and should be considered with care when calculations are restricted 
to axially symmetric shapes
 ~\cite{bonatsos,Rodrigues,Robledo}. 
 Our calculations clearly show that the transitional region of the N=90 isotones is well explained using two and 
three dimensional potential 
energy surfaces within the cranking approximation. 
The quadrupole moments, the moments-of-inertia and the calculated  
$R_{4/2}$ ratio all
agree very well with experiment. The impact of our study for the characterization of phase transition in nuclei needs to be further elucidated. 
It may also indicate a restricted validity of the concept 
due to the presence of large fluctuations and other degrees of freedom.

\title{Acknowledgements}
This project is supported by the Turkish Automic Energy Authority (TAEK) 
under Project No. DPT-04K120100-4.

\end{document}